%% file: VQE-2-particle.tex
\documentclass[showpacs,showkeys,nofootinbib,amsfonts,amssymb,floatfix
,twocolumn,tightenlines
,groupaddresses
]{revtex4-2}
\pdfoutput=1

\usepackage{graphicx}  
\usepackage{mathtools}
\usepackage{amsmath, amssymb, bm,bbm}   
\usepackage{xcolor}
\usepackage{natbib}
\usepackage{hyperref}

\usepackage{longtable}
\usepackage{csvsimple} 
\usepackage{adjustbox} 
\usepackage{tabularx} 

\usepackage[normalem]{ulem} 

\usepackage{ifthen}

\usepackage{siunitx} 
\usepackage{balance}

\newcolumntype{C}{>{\centering\arraybackslash}X}

\newcommand*\xbar[1]{%
   \hbox{%
     \vbox{%
       \hrule height 0.5pt 
       \kern0.3ex
       \hbox{%
         \kern-0.1em
         \ensuremath{#1}%
         \kern-0.1em
       }%
     }%
   }%
} 

\begin{document}

\title{
Elastic scattering on a quantum computer}
\author{%
Muhammad Yusf}
\email{mf1478@msstate.edu}

\author{%
Ling Gan}
\email{lg1208@msstate.edu}

\author{%
Cameron Moffat}
\email{cpm264@msstate.edu}

\author{%
Gautam Rupak}
\email{grupak@ccs.msstate.edu}

\affiliation{Department of Physics \& Astronomy and HPC$^2$ Center for 
Computational Sciences, Mississippi State
University, Mississippi State, MS 39762, USA}

\begin{abstract}
Scattering probes the internal structure of quantum systems. We calculate the two-particle elastic scattering phase shift for a short-ranged interaction on a quantum computer. Short-ranged interactions with a large scattering length or shallow bound state describe a universality class that is of interest in atomic, condensed matter, nuclear, and particle physics. The phase shift is calculated by relating the ground state energy of the interacting particles in a harmonic trap. The relaxation method is used as the variational quantum eigensolver for the ground state calculation. 
Schmidt decomposition is used to reduce quantum circuits nominally requiring tens of qubits to 2-qubit circuits, thus  reducing the noise in quantum measurements. Calculations in multi-particle systems with many-body interactions would benefit from this reduction of qubits in noisy quantum processors.
\end{abstract}
\maketitle

\section{Introduction}
\label{sec:Intro}
Scattering---elastic and inelastic---is a central tool for probing the internal structure of a quantum system. The discovery of the atomic electronic structure in Rutherford's 1911 gold-foil scattering experiment and the parton structure of nucleons in SLAC-MIT deep inelastic scattering experiments~\cite{Friedman:1991,Kendall:1991, Taylor:1991} of the late 1960s are well-known examples of that. Scattering experiments describe the time evolution of quantum systems subjected to a Hamiltonian associated with the interactions between the scattering objects. On the other hand, nuclear many-body calculations on classical computers are typically done in Euclidean time using Monte Carlo simulations. Static properties such as the energy levels of atomic nuclei  can be calculated from first principles using microscopic nuclear interactions, see Ref.~\cite{Lu:2018} and references therein. However, working in Euclidean time, the extraction of real-time evolution of quantum systems becomes difficult. Currently, L\"{u}schur's method~\cite{Luscher:1991ux}, the so-called Busch formula~\cite{Busch:1998cey} and the spherical-wall method~\cite{Carlson:1984, Borasoy:2007} are used to extract elastic scattering phase shifts in classical calculations. A general method for classical calculations of inelastic scattering amplitude for nuclear many-body systems in Euclidean-time  has not been developed yet. In relativistic field theory, the Maiaini-Testa theorem~\cite{Maiani:1990} precludes calculations of inelastic cross sections away from kinematical thresholds from Euclidean-time correlation functions.

In contrast to classical computations, quantum computations (QCs) are done in Minkowski-time as unitary evolution of quantum states. Thus, there is an expectation that it will provide an advantage over classical computation in evaluating dynamical properties. Further, Monte Carlo calculations in Euclidean time suffer from the fermionic sign problem~\cite{Loh:1990, Aarts:2016, Carlson:2015, Elhatisari:2017} which becomes severe for large systems. This is another area where QCs are expected to be advantageous to classical computations. However, it is known that QC on the  Noisy Intermediate-Scale Quantum (NISQ) era computers has challenges. These include efficient qubitization of the fermionic and bosonic degrees of freedom~\cite{Bauer:2020,Alexandru:2022}, and the extraction of meaningful signals from noisy measurements. Even if we had access to an ideal universal quantum computer, QC of scattering cross sections is not straightforward~\cite{Jordan:2012,Jordan:2014}.

In a recent work~\cite{Bedaque:2022ftd}, a general algorithm for calculating inelastic reactions $a(b,\gamma)c$, $a(b,c)d$ where $a$, $b$, $c$, $d$ are atomic nuclei or particles, and $\gamma$ a photon was presented. The frequency of 
the Rabi oscillations between the forward and backward reactions that would invariably result from the unitary time evolutions were related to the reaction amplitude. The algorithm was demonstrated for a single particle in one spatial dimension. Extension to higher dimensions and with more particles would be needed to demonstrate a quantum advantage. 

In this work, we calculate the two-particle elastic scattering phase shift for a short-ranged interaction. It describes a universal class of interactions of relevance in atomic, condensed matter, nuclear, and particle physics. At momenta that are small compared to the inverse range of the interaction, the two-particle interaction can be treated as zero-ranged in the  leading order approximation. When the strength of the interaction is such that the two-particle scattering length is much larger than the range of the interaction (a resonating system), the low momentum phase shift becomes universal, independent of the details of the interaction. This happens naturally, for example, for two-nucleon and atomic $^4$He-$^4$He $s$-wave 
 scattering~\cite{Grisenti:2000, Wiringa:1995}. Atomic systems near a Feshbach resonance can be tuned to be in this universal regime as well~\cite{Ohara:2002, Regal:2003}. 

The phase shift calculation in 1-spatial dimension uses the Busch formula that measures the energy shift of particles in a harmonic trap due to the interaction. The classical calculation of the three-particle system in a harmonic trap was treated in Ref.~\cite{Luu:2006xv} that has been extended to many-body calculations as well~\cite{Stetcu:2007ms}. Experimental realization of such a quantum system in a harmonic trap has been possible~\cite{Kohl:2005,Stoferle:2006}. 

In section~\ref{sec:Busch}, we present a derivation of the  Busch formula relating the relative two-particle center-of-mass (cm) energy to the scattering phase shift. The lattice Hamiltonian for the energy calculation is presented in section~\ref{sec:VQE}. The eigenenergy calculation of an eigenstate through quantum phase estimation is difficult since that would involve eigenstate preparation. Adiabatic time evolution is a physically motivated approach for state preparation. However, it is computationally expensive to obtain accurate results, even in a small system such as a particle in a box, requiring many small time steps with each time step incurring errors in NISQ machines significant enough to render the calculation impractical. Thus, the energy calculation is performed using a variational quantum eigensolver (VQE) where a trial wave function in the general two-particle coordinate system is iteratively evolved for an upper bound on the ground state energy. 
We use the relaxation method~\cite{Schroeder:2017, Jackson:1998} for the iterations that are done classically. Results are presented in section~\ref{sec:Results}. Calculations using both ideal simulations and a physical quantum processor are presented. 
The conclusions are in section~\ref{sec:Conclusions}.

\section{Phase shift in 1-D}
\label{sec:Busch}

The 1-D Schr{\"o}dinger equation for two particles interacting with a short-ranged force in $s$-wave is
\begin{align}
    -\frac{1}{2\mu}\frac{d^2}{dx^2}\phi(x) + c_0\delta(x)\phi(x)= E\phi(x)\,,
\end{align}
where $x$ is the relative coordinate, $\mu$ the reduced mass, $E$ the relative energy. $c_0>0$ corresponds to a repulsive potential with an outgoing scattering solution $\phi(x>0)= t \exp(i p x)$ where momentum $p=\sqrt{2\mu E}$. For incident wave $\phi(x<0)=\exp(i p x)$, the complex transmission amplitude is 
\begin{align}\label{eq:phaseshift}
    t=\frac{\hat{p}}{1+\hat{p}^2}e^{i\delta(p)}\, ,
\end{align}
where the phase shift $\delta(p) = \arctan(-1/\hat{p})$ with dimensionless $\hat{p}= p/(\mu c_0)$. 

The normalized bound state solution is $\phi_B(x)=\sqrt{\gamma} \exp(-\gamma|x|)$ for $c_0<0$  where the binding energy $B=\gamma^2/(2\mu)$ with $\gamma = -\mu c_0$. Busch et al.~\cite{Busch:1998cey} defines a scattering length $a_0=-1/(\mu c_0)$ that gives $\gamma=1/a_0$ coinciding with the 3-D effective range expansion for short-ranged interactions~\cite{Bethe:1949}.

The Busch formula ~\cite{Busch:1998cey} relating energy shift to scattering phase shift is derived from the general coordinate Schr{\"o}dinger equation where we have in a simple harmonic trap with frequency $\omega$ 
\begin{multline}\label{eq:Hcontinuum}
   \left[ -\frac{1}{2m_1}\frac{d^2}{dx_1^2}-\frac{1}{2m_2}\frac{d^2}{dx_2^2}+
   \frac{1}{2}m_1\omega^2 x_1^2+\frac{1}{2}m_2\omega^2 x_2^2\right.\\
   \left.+ c_0\delta(x_1-x_2)
   \right]\Psi(x_1,x_2)= E_\text{total}\Psi(x_1,x_2)\,,
\end{multline}
for two particles with masses $m_1$ and $m_2$ at locations $x_1$ and $x_2$, respectively. 
In terms of cm $X=(m_1 x_1+m_2 x_2)/(m_1+m_2)$, relative $x=x_1-x_2$ coordinates, we write for the cm motion
\begin{align}
    \left[ -\frac{1}{2M}\frac{d^2}{dX^2}+\frac{1}{2}M \omega^2 X^2\right] \psi_n(X)= (n+\frac{1}{2}) \omega \psi_n(X)\,,
\end{align}
which describes a particle of mass $M=m_1+m_2$ moving in a harmonic trap of frequency $\omega$ with energy eigenfunction $ \psi_n(X)$. The energy spectrum of the cm motion is the usual simple harmonic oscillator (SHO) energies $(n+1/2)\omega$ with $n=0,1,\cdots$. 

The relative motion is governed by the Schr{\"o}dinger equation
\begin{align}\label{eq:E_rel}
  \left[  -\frac{1}{2\mu}\frac{d^2}{dx ^2}+\frac{1}{2}\mu\omega^2x^2 +c_0\delta(x)\right]\phi(x)= E \phi(x)\,,
\end{align}
that is solved exactly in a SHO basis by inserting in Eq.~(\ref{eq:E_rel})
\begin{align}
    \phi(x)=\sum_{n=0}^\infty\phi_n(x)\,,
\end{align}
where $\phi_n(x)$ are the energy egienfunctions of the SHO with mass $\mu$ and frequency $\omega$. A self-consistent treatment~\cite{Busch:1998cey} leads to 
the two-particle Green's function relation
\begin{align}
\sum_{n=0}^\infty\frac{\phi_m(0)\phi^\star_m(0)}{E_m-E}=-\frac{1}{c_0}\,.
\end{align}
The $1/c_0$ term on the right hand side above is related to the phase shift in Eq.~(\ref{eq:phaseshift}). The left hand side can be solved exactly using the known eigenfunctions $\phi_n(x)$ resulting in the master equation
\begin{align}\label{eq:Busch}
    p\cot\delta = -\frac{p^2}{\mu c_0}=\frac{1}{2}\frac{p^2}{\sqrt{\mu\omega}}\frac{\Gamma(\frac{1}{4}-\frac{p^2}{4\mu\omega})}{\Gamma(\frac{3}{4}-\frac{p^2}{4\mu\omega})}\, ,
\end{align}
where only the even $n=0,2,\cdots,$ eigenfunctions $\phi_n(0)$ contribute at the origin. 
$p\cot\delta$ has poles at energies $E=(2n+1/2)\omega$ and zeros at energies $E=(2n+3/2)\omega$, respectively, for $n=0,1,\cdots$. Several limits can be derived from the above that can serve as checks on the numerical results.

\subsection{Weak coupling limit}

In the limit $|c_0|\ll 1$, one can use 1st order perturbation theory to calculate the energy shift:
\begin{multline}
  \Delta E_{2n} = \langle\phi_{2n} |c_0\delta(x)|\phi_{2n}\rangle=c_0|\phi_{2n}(0)|^2\\=\sqrt{\frac{\mu\omega}{\pi}} {n-1/2\choose n}c_0\,,
  =\sqrt{\frac{\mu\omega}{\pi}}\frac{(2n-1)!!}{2^n n!} c_0\,,
\end{multline}
that matches exactly with the weak coupling expansion of Eq.~(\ref{eq:Busch}) with $E\approx (2n+1/2)\omega+\Delta E_{2n}$ for small $\Delta E_{2n}\ll \omega$. We used the notation ${n\choose p}$ for combinatorial $n$ choose $p$. The weak coupling limit is more precisely defined as $c_0\ll\sqrt{\omega/\mu}$. 

\subsection{Strong coupling limit}

We can provide a straightforward verification of the strong coupling limit $|c_0|\gg \sqrt{\omega/\mu}$ as a 1st order perturbation in the case of an attractive interaction:
\begin{align}
    \Delta E = \langle\phi_B|\frac{1}{2}\mu\omega^2 x^2|\phi_B\rangle=\frac{\mu\omega^2}{4\gamma^2}\,,
\end{align}
where the normalized bound state wave function $\langle x|\phi_B\rangle=\sqrt{\gamma}\exp(-\gamma|x|)$ with binding energy $B=\gamma^2/(2\mu)$. Directly from Eq.~(\ref{eq:Busch}), expanding in $\omega\ll E$
\begin{align}
    &-\mu c_0\approx2\sqrt{\mu\omega}\sqrt{-\frac{E}{2\omega}}\left[1-\frac{1}{8}{1/2\choose2}\left(-\frac{2\omega}{E}\right)^2\right]\,,\nonumber\\
    \Rightarrow &E+B=\Delta E\approx\frac{\mu\omega^2}{4\gamma^2}\,,
\end{align}
in agreement with the perturbative result 
where we used $\gamma = -\mu c_0$.

\subsection{Energy shifts}
\label{subsec:Eshift}
\begin{figure}[tbh]
\begin{center}
\includegraphics[width=0.47\textwidth,clip=true]{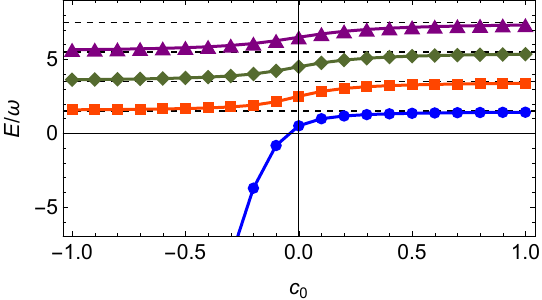}
\end{center}
\caption{\protect The (blue) circles, (red) squares, (green) diamonds and (purple) triangles are exact numerical calculations of the relative energy $E$ for various couplings $c_0$ in the continuum and infinite volume. The colored solid curves are just to guide the eye. The (black) dashed horizontal lines identify the poles in Eq.~(\ref{eq:BuschB}) at $E=(2n+3/2)\omega$. }
\label{fig:energyShift}
\end{figure}

The qualitative nature of the energy shift for nonzero $c_0$ can be understood from Eq.~(\ref{eq:Busch}),  
written in an alternate form
\begin{align}\label{eq:BuschB}
   -c_0=2\sqrt{\frac{\omega}{\mu}} \frac{\Gamma(\frac{3}{4}-\frac{E}{2\omega})}{\Gamma(\frac{1}{4}-\frac{E}{2\omega})}\,,
\end{align}
that has zeros at $E/\omega=2n+1/2$ and poles at $E/\omega=2n+3/2$ for non-negative integer $n$. As the RHS goes through zero (where $c_0=0$), it changes sign while approaching the poles (where $c_0=\pm\infty$) on either side of the zeros. One finds~\cite{Busch:1998cey}, for repulsive interactions $c_0>0$, solutions between the zeros and the poles for $\frac{1}{2}<E/\omega<\frac{3}{2}$, $\frac{5}{2}<E/\omega<\frac{7}{2}, \cdots$. For attractive interactions $c_0<0$, the solutions originate at the poles and terminate at the zeros for $E/\omega<\frac{1}{2}$, $\frac{3}{2}<E/\omega<\frac{5}{2}, \cdots$, see Fig.~\ref{fig:energyShift}.   

The energy shifts in the presence of the two-body interaction are numerically calculated in Fig.~\ref{fig:energyShift}. At $c_0=0$ the energy curves go through $ E=(2n+1/2)\omega$ for $n=0$ $1,\cdots$.  We use $m_1=\SI{940}{\mega\eV}=m_2$ and $\omega=\SI{10}{\mega\eV}$. At any coupling strength, numerically solving for $E$ allows one to calculate the scattering phase shift using Eq.~(\ref{eq:Busch}). The energy calculation is described in the next section.

\section{Variational Quantum Eigensolver}
\label{sec:VQE}
The phase shifts can be calculated once the relative eigenenergies $E$ at a fixed coupling $c_0$ are known. As we work in general coordinates, the identification of a generic eigenstate gets difficult since one has to add the cm energy $E_\text{cm}=(n+1/2)\omega$ to the relative energies shown in Fig.~\ref{fig:energyShift}. However, the ground state can always be identified as the lowest energy state. Thus we calculate  the total energy $E_\text{total}$ of the ground state and then obtain the relative energy $E=E_\text{total}-\omega/2$ by subtracting the lowest cm energy. To get phase shifts at different momenta, we simply vary the SHO frequency $\omega$ with a larger value resulting in a more tightly bound wave function and a higher momentum phase shift. 

The discretized Hamiltonian, in a box with $L$ sites, in the 2nd-quantization language is written as 
\begin{multline}\label{eq:Hlattice}
\hat{H}=-\hat{h}\sum_{\sigma=1}^2\sum_{l=0}^{L-2}[\hat{\psi}_{\sigma,l+1}^\dagger \hat{\psi}_{\sigma,l}+\operatorname{h.c.}]\\
+\frac{1}{2} \hat{m}\hat{\omega}^2\sum_{\sigma=1}^2\sum_{l=0}^{L-1}[l-\frac{L-1}{2}]^2\hat\psi_{\sigma,l}^\dagger \hat\psi_{\sigma,l}\\
+c_0(\hat\psi^\dagger_{2,l}\hat\psi_{2,l})(\hat\psi^\dagger_{1,l}\hat\psi_{1,l})\,,
\end{multline}
where the three terms are the kinetic energy, single-particle harmonic potential and two-particle interactions, respectively. 
Dimensionful quantities in units of the lattice spacing $b$  are indicated with a hat. The physical extent of the box is defined from $-(L-1) b/2$ to $(L-1) b/2$ for odd integer $L$.  We define the hopping parameter $\hat h=1/(2\hat m)$ with $m_1=m=m_2$. The diagonal term $2 \hat h\hat\psi_{\sigma,l}^\dagger\hat \psi_{\sigma,l}$ in the kinetic operator at each lattice site $l=0,1,\cdots, L-1$ for particle type $\sigma=1, 2$ is not shown. 

VQE is implemented through the relaxation method~\cite{Schroeder:2017, Jackson:1998}. The differential operator in the non-relativistic Hamiltonian is a Laplacian whose solutions are harmonic functions. These have the property that the function at the center of a region is given by the average of the function on the boundary of the region. This is the underlying principle of the relaxation method. Given some trial wave function in 1st quantized form $\Psi(x_1,x_2)$ we write an updated 
wave function~\cite{Schroeder:2017} on the lattice from Eq.~(\ref{eq:Hcontinuum})
\begin{align}\label{eq:relax}
\hat{\Psi}^{(\text{new})}_{i_1,i_2}&=\frac{4 \hat{h} \hat{\xbar{\Psi}}_{i_1,i_2}}{4 \hat{h} +[\hat{V}_{i_1,i_2}-\hat{E}_\text{total}]\hat{\Psi}_{i_1,i_2}]}\,,\nonumber\\
\hat{V}_{i_1,i_2}&=\frac{1}{2} \hat m \hat\omega^2(i_1^2+i_2^2) + c_0 \delta_{i_1, i_2}\,,\nonumber\\
 \hat{\xbar{\Psi}}_{i_1,i_2} &=\frac{\Psi_{i_1+1,i_2}+\Psi_{i_1-1,i_2}+\Psi_{i_1,i_2+1}+\Psi_{i_1,i_2-1}}{4}\,,
\end{align}
where the updated wave function is proportional to the average of the nearest neighbors. $\Psi_{i_1, i_2}$ is the two-particle wave function on the lattice with particle 1 at site $i_1$ and particle 2 at site $i_2$. The relaxation method is the classical component of the VQE algorithm once the energy $\hat{E}_\text{total}$ is calculated on the quantum simulator. In the Gauss-Siedel method outlined in Ref.~\cite{Schroeder:2017}, only a single quantum computation of $\langle H\rangle$ with the trial wave function is needed at the first step in the iterations, with the rest of the updates done classically. We also implemented the Jacobi method where each iteration requires a quantum computation of $\langle H\rangle$ with the updated wave function. It is computationally expensive as each iteration requires a QC. On noisy simulators, we need around 30 iterations to get a ground state estimate that is within a few percent of the exact value. We do not use the Jacobi method for our final result. 

The particular implementation of VQE we employ is different from typical applications where the trial wave function ansatz is parameterized with unitary rotation angles that are varied to minimize the ground state energy calculation.
We find optimization with initial angles chosen to reproduce the trial functions used in section~\ref{sec:Results} gives results with errors of about 20\% for $L=5$ even on an ideal simulator~\cite{Moffat:2024}. The error is found to be larger as $L$ increases. We expect the errors to be larger on physical devices due to the large number of two-qubit entangling gates required in such an approach. The states parameterized with unitary rotations often include superposition of states other than those with a single particle of each type as intended.
Instead, in our calculations, we treat the wave function amplitudes in position space as the parameters that are varied according to Eq.~(\ref{eq:relax}) for energy minimization. The evolved states in the relaxation method remain within the subspace of states with two particles, one of each kind.

\subsection{Energy expectation and Schmidt decomposition}

The computation of $\hat{E}_\text{total}$ involves expectation value $\langle\hat{H}\rangle$ calculations on a quantum computer. In a box with $L$ lattice sites, we require $2L$ qubits to represent a two-particle state. The fermionic degrees of freedom in the Hamiltonian, Eq.~(\ref{eq:Hlattice}), are represented by Pauli matrices using the Jordan-Wigner transformation~\cite{Jordan:1928}. The kinetic term in lattice units is
\begin{multline}\label{eq:KE}
\widehat{\text{KE}}
=4 \hat{h}-\frac{\hat{h}}{2}\sum_{l=0}^{L-2}[ X_{l+1} X_l+ Y_{l+1}Y_{l}\\
+X_{l+1+L} X_{l+L}+Y_{l+1+L}Y_{l+L}]\,,
\end{multline}
where we put the particles of type-1 (the 1st particle) on the lowest $L$ qubits, and the particles of type-2 (the 2nd particle) on the highest $L$ qubits. The two-particle coordinate $(i,j)$ for particles type-1 and type-2, respectively, is mapped to the computational basis $|2^{j+L}+2^i\rangle$. This is done so that hopping of the particles is to the nearest sites to avoid any Jordan strings. We adopt the notation that only the non-identity operations in terms of the Pauli matrices are shown explicitly. The diagonal term $2 \hat h\hat\psi_{\sigma,l}^\dagger\hat \psi_{\sigma,l}$ contributes an energy of $4 h$ at each lattice site shifting the overall energy by $4h$ that we do not  calculate explicitly on the quantum computer. The harmonic trap contribution to the potential energy in Eq.~(\ref{eq:Hlattice}) uses the same qubit ordering of the two types of particles as the $\widehat{\text{KE}}$ term, and gives in lattice units 
\begin{align}\label{eq:PESHO}
\widehat{\text{PE}}_\text{SHO}=&\frac{1}{4}\hat{m}\hat{\omega}^2\sum_{l=0}^{L-1}[l-\frac{L-1}{2}]^2
[\mathbbm{1}_l - Z_l]\nonumber \\
&+\frac{1}{4}\hat{m}\hat{\omega}^2\sum_{l=0}^{L-1}[l-\frac{L-1}{2}]^2[\mathbbm{1}_{l+L} - Z_{l+L}] \nonumber\\
=&\frac{ \hat m\hat\omega^2 L(L^2-1)}{24}\nonumber\\
&-\frac{\hat{m}\hat{\omega}^2}{4}\sum_{l=0}^{2L-1}[
\operatorname{mod}(l,L)-\frac{L-1}{2}]^2 Z_l\,.
\end{align}
The two particle interaction at each lattice site $l=0,\cdots, L-1$ is written as
\begin{multline}\label{eq:PEint}
4 (\hat\psi^\dagger_{2,l}\hat\psi_{2,l})(\hat\psi^\dagger_{1,l}\hat\psi_{1,l})\rightarrow\mathbbm{1}_{2l+1}\mathbbm{1}_{2l}
-\mathbbm{1}_{2l+1}Z_{2l}\\-Z_{2l+1}\mathbbm{1}_{2l}+Z_{2l+1}Z_{2l}\,,
\end{multline}
where we put particle of type-1 at the even numbered qubits and particle of type-2 at the odd numbered qubits for convenience---coordinate $(i,j)$ is mapped to computational basis $|2^{2j+1}+2^{2i}\rangle$---that would become evident when we talk about Schmidt decomposition later. 
This gives for the two-particle interaction potential in lattice units
\begin{align}
\widehat{\text{PE}}_\text{int.}&= \frac{c_0}{4} L + \frac{c_0}{4} \sum_{l=0}^{L-1} [- Z_{2l} - Z_{2l+1}
+Z_{2l+1} Z_{2l}]\nonumber\\
&=\frac{c_0}{4} L-\frac{c_0}{4}\sum_{l=0}^{2L-1}Z_l
+\frac{c_0}{4} \sum_{l=0}^{L-1} Z_{2l+1} Z_{2l}\,.
\end{align}

We combine the potential energy contributions for convenience into single $Z_l$ and double $Z_{2l+1} Z_l$ terms as 
\begin{align}
    \widehat{\text{PE}}_{Z}  
    =&\frac{ \hat m\hat\omega^2 L(L^2-1)}{24}\nonumber\\
&-\frac{\hat{m}\hat{\omega}^2}{4}\sum_{l=0}^{2L-1}\left\{[
\operatorname{mod}(l,L)-\frac{L-1}{2}]^2-\frac{c_0}{4}\right\} Z_l\,,\nonumber\\
\widehat{\text{PE}}_{ZZ}=&
\frac{c_0}{4} L
+\frac{c_0}{4} \sum_{l=0}^{L-1} Z_{2l+1} Z_{2l}\,.
\end{align}
The Hamiltonian is given by the sum $\widehat{\text{KE}}+ \widehat{\text{PE}}_{Z}+\widehat{\text{PE}}_{ZZ}$. 

The calculation of the energy expectation value $\langle\hat{H}\rangle$  is reduced in VQE to the expectation values of Pauli matrices for some trial state 
\begin{align}
|\psi\rangle =\sum_{k=0}^{2^{2L}-1}\psi_{k}|k\rangle\,,
\end{align}
written in the computational basis $|k\rangle$  for $2L$ qubits. 

Consider the measurement of $\langle Z_{2L-1}\rangle= \sum_{k,k'}\psi_{k'}^\star\psi_{k} \langle k'|Z_{2L-1}|k\rangle$ for the SHO PE term. A little deliberation leads one to perform the quantum calculation in the reduced Hilbert space of only the highest 
$(2L-1)$th qubit after tracing over the lower qubits where the expectation value is simply one. Specifically, we bipartate the Hilbert space into tensor product  $A\otimes B$ of dimensions $A$ and $B$, respectively. Here, $A$ is a 
 2-dimensional space represented by the highest qubit $q_{2L-1}$ and $B$ a $2^{2L-1}$-dimensional space represented by the lowest qubits $q_{2L-2}\cdots q_0$ with orthonormal bases $|a\rangle$ and $|b\rangle$, respectively. Then one writes~\cite{Book:McIntyre}, including explicit identity operator in the lowest $2L-2$ qubits  
\begin{multline}\label{eq:density}
  \langle\psi| Z_{2L-1}\otimes \mathbbm{1}_B |\psi\rangle = \sum_{a,b}
   \langle\psi| Z_{2L-1}\otimes \mathbbm{1}_B|a,b\rangle\langle a,b|\psi\rangle\\
   =\sum_{a,b}\sum_{a',b'}\langle a,b|\psi\rangle\langle\psi|a'b'\rangle\langle a' b'|Z_{2L-1}\otimes \mathbbm{1}_B |a,b\rangle\\
   = \sum_{a, a',b, b'}[\rho]_{ab;a'b'}[Z_{2L-1}]_{a' a}\delta_{b' b}\\
   =\sum_{a,a', b}[\rho]_{ab;a'b}[Z_{2L-1}]_{a' a}\,,
\end{multline}
where we define the tensor product state $|a,b\rangle\equiv|a\rangle\otimes|b\rangle$ and the density matrix $\rho\equiv |\psi\rangle\langle\psi|$ using standard notations. The relation in Eq.~(\ref{eq:density}) is simplified by marginalizing over the index $b$  where the expectation value of $\mathbbm{1}_B$ in the Hilbert space $B$ is trivial:
\begin{align}
     \langle\psi| Z_{2L-1}\otimes \mathbbm{1}_B |\psi\rangle&=
     \operatorname{Tr}(\rho_A Z_{2L-1})\,\nonumber\\
     [\rho_A]_{aa'}&\equiv \sum_{b}[\rho]_{ab;a'b}\,.
\end{align}

The calculation of the reduced density matrix $\rho_A$ is simplified in the Schmidt decomposition. In the bipartite state
\begin{align}\label{eq:matrixM}
    |\psi\rangle=\sum_{a,b} M_{ab}|a,b\rangle\,,
\end{align}
the probability amplitude $M_{ab}$ is interpreted as a matrix with row-column indices $a$-$b$. Then singular value decomposition (SVD) gives 
\begin{align}
     M=U S V^\dagger\,.
\end{align}
In our calculations we have size $A<B$, and $U$ is of size $A\times A$ and $V^\dagger$ is of size $A\times B$ with a diagonal matrix $S_{aa}=\lambda_a$ of size $A\times A$. The Schmidt decomposition of the state $|\psi\rangle$ is in terms of a new set of bases $|i\rangle_A=\sum_a U_{ai}|a\rangle$ and $|i\rangle_B=\sum_bV_{bi}^\star|b\rangle$ such that
\begin{align}
    |\psi\rangle=\sum_{i=1}^{r\leq A} \lambda_i |i\rangle_A\otimes |i\rangle_B\,,
\end{align}
where $r$ is the Schmidt rank identifying the number of non-zero singular values.  
One notes that the same number of bases from Hilbert space $A$ and $B$ are needed. Even though the Hilbert space of $B$ is much larger than $A$ in our calculations, Schmidt rank $r \leq A$ contains all the entanglement information between the two Hilbert spaces. The reduced density matrix can now be written by tracing over the subspace $B$ as
\begin{align}
    \rho_A=
    \sum_{i=1}^{r\leq A}{}_B\langle i|\psi\rangle\langle\psi|i\rangle_B 
    \doteq\sum_{i=1}^{r\leq A} \lambda_i^2|i\rangle_A {}_A\langle i|\,,
\end{align}
which gives the required expectation value for a generic operator
\begin{align}
     \langle\psi| \mathcal{O}_A\otimes \mathbbm{1}_B |\psi\rangle&=
     \operatorname{Tr}(\rho_A  \mathcal O_A) = \sum_{i=1}^{r\leq A}\lambda_i^2
     {}_A\langle i| \mathcal O_A|i\rangle_A\,.
\end{align}
In the quantum circuit, we prepare a superposition state defined on the reduced space $A$:
\begin{align}
    |\psi_A\rangle =\sum_{i=1}^{r\leq A}\lambda_i |i\rangle_A\,,
\end{align}
which results in the required expectation values. 

Single Pauli expectations such as $\langle Z_{2L-1}\rangle$ are done in the Hilbert space of size $A=2$, represented by a single qubit. To construct the matrix $M$ in Eq.~(\ref{eq:matrixM}) for Schmidt decomposition, we simply divide the qubit state $|\psi\rangle$ in two equal parts, and these become the two rows of the matrix $M$. For generic measurement  $\langle Z_{l}\rangle$ with $l\neq 2L-1$, we simply re-index our site labels for the wave function to put the relevant $q_l$ as the highest qubit, and then the same procedure  applies. 

For two Pauli expectations such as $\langle Z_{2L-1} Z_{2L-2}\rangle$ in Eq.~(\ref{eq:PEint}), we divide the state $|\psi\rangle$ into 4 equal parts, which form the 4 rows of the matrix $M$ in Eq.~(\ref{eq:matrixM}) for Schmidt decomposition. It is evident, we want to put the Pauli matrices whose expectation value  we wish to calculate at the highest two qubits. This explains the ordering of the two-particle  qubits we used for the kinetic and the interaction terms in Eqs.~(\ref{eq:KE}), ~(\ref{eq:PEint}).  

We find that using the Schmidt decomposition to trace out the spectator space $B$ that only involves the identity operator $\mathbbm{1}_B$, gives more accurate results in the noisy QCs. The Schmidt decomposition can be optimized further by noting that one could perform SVD on the smaller square matrix $M M^\dagger$ of size $A\times A$ such that
\begin{align}
    M M^\dagger = U S^2 U^\dagger\,,
\end{align}
which avoids storing and manipulating large matrices with at least one of the dimensions as large as $2^{2L-1}$.  We only need $S$, $U$ to  construct the reduced state $|\psi_A\rangle$.

\section{Results}
\label{sec:Results}

For the numerical work, we continue from subsection~\ref{subsec:Eshift} with particle masses $m_1=\SI{940}{\mega\eV}=m_2$  motivated by the proton-neutron masses. The range of the nuclear potential is set by the Compton wavelength of the pion, and it is roughly around a fermi. At distances much larger than the pion Compton wavelength the nuclear interaction can be treated as point-like. Thus, to make the numerical work physically motivated, we take a lattice spacing $b\sim\SI{1}{\femto\m}$ and $L b\sim \SI{10}{\femto\m}$ such that we can calculate the two-particle scattering phase shifts at momenta $\frac{1}{L b}\ll p\ll \frac{1}{b}$ that approximates the continuum and infinite volume results.

We performed several QCs using the ideal simulator on qiskit~\cite{Qiskit} and the IBMQ-Brisbane (\verb|ibmq_brisbane|) quantum processor unit (QPU) at different  values of box size $L$, lattice spacing $b$, and oscillator frequency $\omega$. The two-particle interaction strength $c_0$ was also varied in the simulator but for the QPU calculations, we limited the measurements to a single, conveniently chosen, non-perturbative $c_0=0.2\gg\sqrt{\pi\omega/(4\mu)}$ value to fully utilize the resources available in the IBMQ Open Plan. 

In Table~\ref{table:energyIBMQB}, the theory parameters $L$, $b$, $c_0$, $\omega$ are listed to identify the specific ``runs" of QCs. The trial wave function was chosen to be the normalized ground state of two free particles in a box that is non-zero only at most 7 lattice sites in the middle of the harmonic trap. We get similar results using the ground state of two non-interacting particles in a harmonic trap.  Using a trial wave function with a compact support makes the SVD calculation more accurate and gives less noisy results when tested on the noisy qiskit simulator~\cite{Qiskit}.
In Run 1, the full circuit with 10-qubits for $L=5$ is compared with the results from measurements with 1- and 2-qubit circuits that result after SVD. We write the generic 2-qubit circuits using a single entangling gate as explained in appendix~\ref{sec:statepreparation}. The number of measurements (shots) is as indicated, and should result in very small statistical errors scaling as $1/\sqrt{\text{shots}}$. One can see that the SVD circuit measurement is more accurate than the full circuit measurement on QPU, as expected. For Runs 1 and 4, we did not specify any particular initial layout for the physical qubits, and let qiskit select the necessary connected qubits to implement the appropriate circuits.  Runs 2 and 3 use the same parameter values but are run on different physical qubits as indicated in Table~\ref{table:expectationsB}. We selected the qubits that had low posted readout errors at runtime. The results are noticeably different. Coincidentally, combining the 1-qubit measurements from Run 2 with the 2-qubit measurements from Run 3 gives nearly perfect results in this particular instance. 

Looking more closely at the discrepancy between the Run 2 and Run 3 measurements in Table~\ref{table:expectationsB}, one notices that the individual expectation value measurements of $\Delta\text{KE}$, $\Delta\text{PE}_{\text{ZZ}}$ and $\Delta\text{PE}_{\text{Z}}$ are similar, differing no more than 10\% between the runs. However, there are large cancellations between the various terms in the total energy calculation. This coincidentally gives   $E_{\text{total}}=\SI{22.5}{\mega\eV}$ when 
using $\Delta\text{PE}_{\text{Z}}$ from Run 2 with $\Delta\text{KE}$ and $\Delta\text{PE}_{\text{ZZ}}$ from Run 3 in close agreement with the exact result $\SI{22.33}{\mega\eV}$ for the trial wave function, see Table~\ref{table:energyIBMQB}.
Nevertheless, we use the noisy results as listed in Table~\ref{table:energyIBMQB} in the relaxation method for the ground state calculation to evaluate how well the relaxation method performs in calculating the ground state energy.

\begin{table*}[htb]
\centering
\caption{On IBMQ-Brisbane, the single Pauli measurements were done with 10000 shots and the double Pauli measurements were done with 20000 shots, with 
\texttt{optimization\_level=2}. The last four columns are energy measurements relative to the exact energy.}
\begin{ruledtabular}
\begin{tabular}{rrlrlccccc}
Run &$L$ & $b$ (fm) & $\omega$ (MeV)& $c_0$ & Exact (MeV)& Full ideal &Full QPU
& SVD  ideal &SVD QPU
\\ \hline \rule{0pt}{0.9\normalbaselineskip}
\begin{filecontents*}{IBMQ_Runs_c0p20.csv}
rr,L,a,w,cc,exact,fullideal,fullqpu,svdideal,svdqpu
1,5,1.5,5,0.2,21.33,1.003,1.873,0.9599,1.009
2,13,0.8,10,0.2,22.33,9999,9999,0.9905,1.459
3,13,0.8,10,0.2,22.33,9999,9999,0.9905,1.346
4,15,1.0,6,0.2,15.52,9999,9999,0.9723,1.324
\end{filecontents*}
\csvreader[head to column names, late after line=\\]{IBMQ_Runs_c0p20.csv}{}
{\  $\rr$ & \L & \a &\w &\cc
& \ifthenelse{\equal{\exact}{9999}}{---}{\num[parse-numbers=false]{\exact}}
& \ifthenelse{\equal{\fullideal}{9999}}{---}{\num[parse-numbers=false]{\fullideal}}
& \ifthenelse{\equal{\fullqpu}{9999}}{---}{\num[parse-numbers=false]{\fullqpu}}
& \ifthenelse{\equal{\svdideal}{9999}}{---}{\num[parse-numbers=false]{\svdideal}}
&\ifthenelse{\equal{\svdqpu}{9999}}{---}{\num[parse-numbers=false]{\svdqpu}}
}
\end{tabular}
\end{ruledtabular}
 \label{table:energyIBMQB}
\end{table*}

\begin{table*}[htb]
\centering
\caption{ Expectation components  $\Delta\text{KE}=\text{KE}-4 h$, $\Delta\text{PE}_{ZZ}=\text{PE}_{ZZ}-\frac{c_0L}{4b}$, $\Delta\text{PE}_{Z}=\text{PE}_{Z}-\frac{  m\omega^2 b^2 L(L^2-1)}{24}$. }
\squeezetable
\begin{ruledtabular}
\begin{tabular}{cccccccccc}
\rule{0pt}{0.9\normalbaselineskip}
run & $L$ &qubits & $\Delta\text{KE}$ (MeV)& $\text{KE}$ (MeV) &$\Delta\text{PE}_{ZZ}\text{ (MeV)}$ 
& $\text{PE}_{ZZ}$ (MeV) &
$\Delta\text{PE}_{Z}$ (MeV)& $\text{PE}_Z$ (MeV) & $ E_{\text{total}}$ (MeV)
\\ \hline \rule{0pt}{0.9\normalbaselineskip}
\begin{filecontents*}{expectations_c0p20.csv}
run,L,qbita,qbitb,xxyy,ke,zz,vint,z,sho,energy
2,13,8,9,-111,18.8,118,278,-405,-264,32.6
3,13,84,85,-116,13.5,113,273,-397,-257,30.1
\end{filecontents*}
\csvreader[head to column names, late after line=\\]{expectations_c0p20.csv}{}
{\  \run &\L &[\qbita,\qbitb]
& \ifthenelse{\equal{\xxyy}{9999}}{---}{\xxyy}
& \ifthenelse{\equal{\ke}{9999}}{---}{\ke}
& \ifthenelse{\equal{\zz}{9999}}{---}{\zz}
& \ifthenelse{\equal{\vint}{9999}}{---}{\vint}
&\ifthenelse{\equal{\z}{9999}}{---}{\z}
&\ifthenelse{\equal{\sho}{9999}}{---}{\sho}
& \ifthenelse{\equal{\energy}{9999}}{---}{\energy}
}
\end{tabular}
\end{ruledtabular}
 \label{table:expectationsB}
\end{table*}  

\begin{figure}[tbh]
\begin{center}
\includegraphics[width=0.47\textwidth,clip=true]{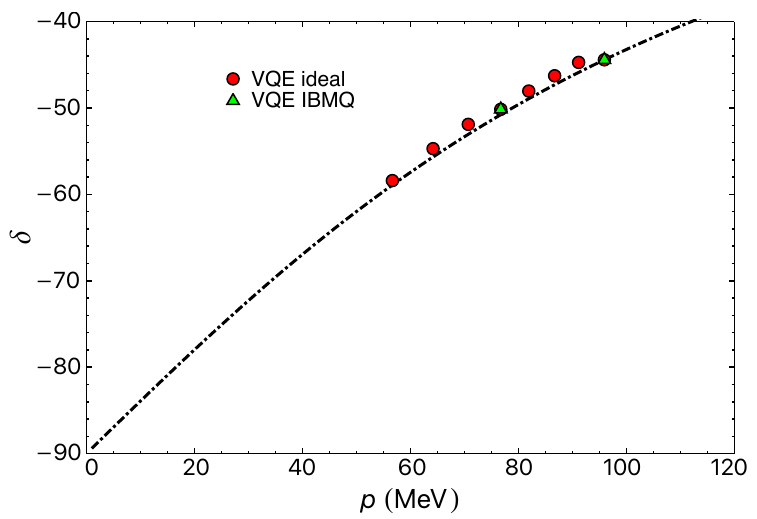}
\end{center}
\caption{\protect Two-body phase shift data calculated on the qasm-simulator as explained on the text. These are compared with the analytical continuum, infinite volume (black) dot-dashed curve.}
\label{fig:phaseshiftSimulator}
\end{figure}

 In Fig.~\ref{fig:phaseshiftSimulator} we show the phase shift calculated using VQE with the Gauss-Siedel relaxation method on an ideal simulator and QPU measurements from Table~\ref{table:energyIBMQB}.   We performed 200 iterations on a classical computer though around 30 iterations would have been sufficient to get a variational ground state energy within a few percent of the exact result. 
 The lowest three momentum data points were calculated at lattice spacing $b=\SI{1.25}{\femto\m}$, lattice size $L=17$ where the SHO oscillator frequency was varied from $\omega=\SI{3}{\mega\eV}$ to  $\SI{5}{\mega\eV}$ in $\SI{1}{\mega\eV}$ interval. The Hilbert space for this lattice size is $2^{34}\approx 2\times 10^{10}$ dimensional, yet, we calculate the expectation values using at most $2^2$ dimensional mixed state after Schmidt decomposition.
 
 The next 4 data points, in Fig.~\ref{fig:phaseshiftSimulator}, were at $b=\SI{1.0}{\femto\m}$, $L=15$ and  from $\omega=\SI{6}{\mega\eV}$ to $\SI{9}{\mega\eV}$ in $\SI{1}{\mega\eV}$ interval. The highest momentum point was at  $b=\SI{0.8}{\femto\m}$, $L=13$ and $\omega=\SI{10}{\mega\eV}$. The lattice parameter $b$, $L$ choices give phase shift quite comparable to the continuum, infinite volume
  black dot-dashed curve. The QPU results are on top of the ideal simulator results. The Gauss-Siedel relaxation method is very efficient in reproducing the ground state wave function and energy. This leads to a perfect match to the ideal VQE results. 

The small discrepancy between the analytical and numerical results in Fig.~\ref{fig:phaseshiftSimulator} results from lattice artifacts which can be removed using smaller lattice spacings and larger boxes.  Larger boxes increase the number of circuits. We chose box sizes and lattice spacings in accordance with the available computational resources while keeping the errors within a few percent.

Before concluding, we comment on a few recent works on scattering on quantum computers. The work by Sharma et al.~\cite{Sharma:2023} calculates the two-particle scattering phase shift as well. They work in relative coordinates where a hard spherical-wall boundary condition is imposed, and extract the phase shift from the excited state energies. The excited states above the bound ground state are the discrete lattice spectrum of the continuum states.  Their calculations were done in an abstract energy eigenbasis such that $n$ qubits are required to calculate phase shifts at $(n-1)$ momenta from the $(n-1)$ excited states above the ground state. The analytic phase shift is reproduced for $n\leq 4$ but for larger qubits the noise in the quantum measurements  renders the calculation inaccurate. A hybrid approach for two-nucleon scattering where the spin degrees of freedom are evolved on a quantum computer and the spatial wave function is evolved on a classical computer was presented in Ref.~\cite{Turro:2023dhg}. They calculated the trajectories for two-nucleon scattering though a phase shift was not evaluated for us to compare with. 

During the writing of this manuscript, we discovered a recent work by Wang et al.~\cite{Wang:2024} that also proposes using a harmonic trap for phase shift calculations. In their work, they use the rodeo algorithm~\cite{Choi:2021} to calculate the energy eigen values for the phase shift determinations. Two example calculations using a spherical-well potential and a Woods-Saxon potential were performed on an ideal qiskit simulator.  QC on a physical device was not considered~\cite{Wang:2024}.

\section{Conclusions}
\label{sec:Conclusions}

We calculated the two-particle scattering phase shift for a short-ranged interaction on a quantum computer. The 
calculation used the Busch formula~\cite{Busch:1998cey} to relate the relative energy of two interacting particles in a harmonic trap to the phase shift. As the exact form of the Busch formula is known, the challenge is to calculate the energies for the various eigenstates of the system. 

The ground state energy was used for the phase shift calculations. It was obtained  using a VQE where the probability amplitudes of the trial wave function in coordinate space were used as the variational parameters. For two particles, this required  $L^2$ amplitudes on a lattice with $L$ sites. In the variational calculations, we used the relaxation method~\cite{Schroeder:2017,Jackson:1998} which requires initial inputs of the trial wave function and the corresponding energy that was calculated on a quantum computer.

The Hamiltonian for the energy expectation values was defined in a $2^{2L}$ dimensional Hilbert space in 2nd quantization---a qubit per lattice site for each type of particles. For the two-particle system, the VQE requires expectation value measurements  of at most two Pauli matrices at a time that we reduced to measurements on two-qubit circuits instead of on 2L-qubit circuits  using Schmidt decomposition. The two-qubit circuits needed only a single CNOT gate thus reducing the noise on NISQ hardware measurements compared to the same calculation on 2L qubits that didn't use the Schmidt decomposition. 
We found the variational calculations to be robust against  noisy measurements. For the system we studied, not only the ground state energy but also the  ground state wave function was obtained  
accurately.

Generalization to many-body systems with $A$ particles would require $A\times L$ qubits on a lattice with $L$ sites.
In the relaxation method, one would need $L^A$ amplitudes even though the Hilbert space grows as $2^{AL}$. 
A rough estimate of the memory requirement for the classical component of the hybrid calculation would restrict us to about 10 particles in one spatial dimension or a few particles in three spatial dimensions with $L\sim10$ on a laptop. However, these are not hard limits as one can work sequentially with the probability amplitudes in Eq.~(\ref{eq:relax}) from numbers stored on a hard drive. Thus, the relaxation method might remain practical for larger systems. Schmidt decomposition can be applied to reduce the number of qubits on the quantum measurements. A $n$-body interaction would be reduced to measurements on $n$-qubit circuits instead of the full  $(A \times L)$-qubit circuit. 
The reduction in qubit utilization was shown to reduce the noise in measurements. 
This would be beneficial in many-body calculations. There is usually a hierarchy where the lower-body interactions dominate at low density. 
 For example, in nuclear systems, chiral symmetry limits the interactions to at most 4-nucleon interactions at the leading order approximation~\cite{Yang:2021vxa}.

Future work should involve calculations in higher dimensions. The current calculation involving a delta function should be extended to include derivatives of delta function which would bring in the effective range correction to the phase shift. Moreover, few particles in higher dimensions should be investigated to explore the effectiveness of the  relaxation method given that conventional VQE with unitary rotations was found to have insufficient accuracy for the two-particle system in one dimension.

\acknowledgments
The authors benefited from discussions with Paulo Bedaque, Thomas Cohen, and Ratna Khadka. We thank  Dean Lee for his insightful comments on the manuscript.  
This work was partially supported by grants  DOE DE-SC00211, DOE DE-SC0024286 and NSF PHY-2209184.
Part of the work was completed by GR at the Maryland Center for Fundamental Physics,  Univ. of Maryland, College Park during his sabbatical. 
We acknowledge the use of IBM Quantum services for this work. In this paper we used \verb|ibmq_brisbane|.

\appendix
\section{2-qubit state}
\label{sec:statepreparation}

Consider a generic 2-qubit state
\begin{align}
|\psi\rangle=\alpha_0|00\rangle+\alpha_1|01\rangle+\alpha_2|10\rangle+\alpha_3|11\rangle\,.
\end{align}
If $|\alpha_0|^2+|\alpha_1|^2=0$ or $|\alpha_2|^2+|\alpha_3|^2=0$, then we have product states $|1\rangle\otimes(\alpha_2|0\rangle+\alpha_3|1\rangle)$ or 
$|0\rangle\otimes(\alpha_0|0\rangle+\alpha_1|1\rangle)$, respectively. These  product states can be prepared with 1-qubit gates. In particular the state $a|0\rangle +b |1\rangle$ is created with unitary rotation $R_z[\operatorname{phase}(b)] R_y[2\cos^{-1}( a)]R_z[-\operatorname{phase}(b)]$ for $a>0$ without loss of generality. The generic case not covered by either of the two situations above  is prepared as follows. 

We write the 2-qubit state after Schmidt decomposition as
\begin{multline} \alpha_0|00\rangle+\alpha_1|01\rangle+\alpha_2|10\rangle+\alpha_3|11\rangle\\ = (u\otimes v^\star) (\lambda_1|00\rangle+\lambda_2|11\rangle)\,,
\end{multline}  
where the unitary matrices $u$ and $v^\dagger$ are the left and right matrices used in the SVD. The entangled state $\lambda_1|00\rangle+\lambda_2|11\rangle$ is created from 
$(\lambda_1|0\rangle+\lambda_2|1\rangle)\otimes|0\rangle$ with a single CNOT gate. The $(\lambda_1|0\rangle+\lambda_2|1\rangle)$ state with $\lambda_i \geq 0$ is created with a single $R_y(2\cos^{-1} \lambda_1)$ gate. 
The unitary operations $u$, $v^\ast$ are implemented with single-qubit gates to obtain the desired 2-qubit state. 


\bibliographystyle{apsrev4-1}
\input{VQE-2-particle.bbl}
\end{document}

%% file: VQE-2-particle.bbl
%